\documentclass{ws-mpla}
\usepackage{epstopdf}  
\begin{document}

\markboth{T. DeYoung}
{Neutrino Astronomy with IceCube}

\catchline{}{}{}{}{}

\title{NEUTRINO ASTRONOMY WITH ICECUBE}

\author{\footnotesize TYCE DeYOUNG}

\address{Department of Physics, Pennsylvania State University, \\
104  Davey Laboratory, University Park, Pennsylvania 16802, USA\\
deyoung@phys.psu.edu}

\maketitle

\pub{Received (Day Month Year)}{Revised (Day Month Year)}

\begin{abstract}
  IceCube is a kilometer-scale high energy neutrino telescope under
  construction at the South Pole, a second-generation instrument
  expanding the capabilities of the AMANDA telescope.  The scientific
  portfolio of IceCube includes the detection of neutrinos from
  astrophysical objects such as the sources of the cosmic rays, the
  search for dark matter, and fundamental physics using a very large
  data set of atmospheric neutrinos.  The design and status of IceCube
  are briefly reviewed, followed by a summary of results to date from
  AMANDA and initial IceCube results from the 2007 run, with 22 of a
  planned 86 strings operational.  The new infill array known as Deep
  Core, which will extend IceCube's capabilities to energies as low as
  10 GeV, is also described.

\end{abstract}

\ccode{PACS Nos.: 95.55.Vj, 95.85.Ry, 95.35.+d}

\section{Cosmic Rays, Gamma Rays, and Neutrinos}

The origin of the cosmic rays has been the subject of intense study
since their discovery by Victor Hess in 1912.  Today we know that
cosmic sources accelerate particles to energies above $10^{20}$ eV,
although we are uncertain about the dynamics or indeed the identities
of those sources. The spectrum of the cosmic rays follows a broken
power law; those below a break at approximately 3 PeV known as the
``knee'' are believed to originate from sources in our
Galaxy. Extragalactic origins are inferred for cosmic rays at the
highest energies, as no Galactic objects are  believed powerful
enough to accelerate particles to those energies.  The precise energy
scale of the transition remains a matter of some
debate.\cite{Hillas:2006ms,Blasi:2005wk}

The energetic environments in which high energy cosmic rays are
accelerated are likely to include matter or radiation fields with
which the accelerated hadrons will interact, producing charged pions or
kaons which decay to neutrinos.\footnote{High energy neutrino
  telescopes such as IceCube do not have the ability to distinguish
  the charge of observed particles, so we will not distinguish between
  particles and antiparticles in this review.}  The same processes
will also produce neutral pions, which will decay into gamma rays.
The energies of the gamma rays will be comparable to those of the
neutrinos\cite{Halzen:1997hw}, so sources of very high energy (VHE, used
herein to indicate energies of roughly $100$ GeV $\sim 1$ PeV)
neutrinos would also be candidates for VHE gamma ray emission, and
vice versa.  However, TeV gamma rays may also be produced by
accelerated electrons inverse-Compton
scattering lower energy photons.  Neutrinos would not be produced in this case,
so neutrinos provide an unambiguous indicator of hadronic
acceleration.
Conversely,
neutrino emitters may be ``hidden'' if their local environments are
too dense for the gamma rays to escape,\cite{Berezinsky:1985xp} or may
only be visible at lower ($\sim$10 GeV) energies if the sources are
sufficiently distant that TeV gammas are attenuated by interaction
with the extragalactic background light.\cite{Stecker:1992wi}

Neutrino production through $\pi/K$ decay leads generically to a
neutrino flavor ratio of $ \nu_e$ : $\nu_\mu$ : $\nu_\tau$ = 1 : 2 : 0
at the source.  However, 
for essentially all astrophysically relevant distances and energies,
the neutrinos will oscillate on their way to Earth to an equilibrium
flavor ratio of $1:1:1$.  This prediction of the flavor ratio is not
absolute, however, and deviations can carry information about
astrophysical neutrino
sources.\cite{Rachen:1998fd,Kashti:2005qa,Anchordoqui:2003vc,Razzaque:2005ds}
Very extreme flavor ratios could be indicators of new fundamental
physics, such as neutrino decay.\cite{Beacom:2003nh}

Neutrinos are detected by IceCube when they undergo either charged
current (CC) or neutral current (NC) interactions with nucleons in the
ice.  At VHE energies, these interactions are
deeply inelastic, producing a hadronic shower of relativistic
particles.  In CC interactions, a secondary lepton corresponding to
the flavor of the incident neutrino ($e$, $\mu$, or $\tau$) is also
produced, carrying between one half and three-quarters of the neutrino
energy\cite{Gandhi:1995tf} and aligned with the neutrino direction to
better than a degree at energies above the TeV scale.\cite{Gaisser92}

Muons thus produced may travel through the ice for kilometers in a
nearly straight trajectory, depending on their energy.  They are
detected as long, straight tracks in the detector via the Cherenkov
radiation they emit.  At higher energies, stochastic interactions
(bremsstrahlung etc.) produce small showers that are not resolved
individually but which increase the apparent brightness of the track.
Electrons produce a compact ($\sim$10 m) electromagnetic shower
intermingled with the hadronic shower at the vertex, from which
Cherenkov radiation propagates out in a roughly spherical pattern.
At low energies, tau leptons are indistinguishable from electron
cascades due to the rapid $\tau$ decay.  However, at higher energies,
above the PeV scale, a long tau track can be observed, usually with a
cascade on either end due to the initial neutrino-nucleon interaction
and the final tau decay.\cite{Learned:1994wg}  Events in which only
one cascade is observed within the detector\cite{Beacom:2003nh} or in
which the $\tau$ decays to $\mu$ inside the
detector\cite{DeYoung:2006fg} may also be identifiable.

Large-volume neutrino telescopes such as
IceCube can also be used to search for other particles that would
produce light in the detector, such as magnetic
monopoles, long-lived supersymmetric
particles,\cite{Albuquerque:2003mi} and other exotica.  Searches for such
particles are in progress but are beyond the scope
of this review.

Neutrino detection is complicated by the background of atmospheric
muons produced via meson decay in cosmic ray air showers.  Neutrino
telescopes are sited at least a kilometer below the surface to reduce
this muon flux, but the high-energy tail of the flux penetrating to
these depths is still many orders of magnitude more numerous than the
VHE neutrino fluxes of interest.  This background is controlled by
searching for upward-going muons, which must have been produced in
neutrino interactions below the detector.  Other event signatures
characteristic of neutrinos, such as isolated cascades within the
detector or downgoing muon tracks starting from a neutrino interaction
vertex observed within the detector, can also be used.

Potential sources of VHE neutrinos include Galactic sources such as
supernova remnants and extragalactic sources such as active galactic
nuclei and gamma ray bursts; several recent
reviews\cite{Halzen:2002pg,Boettcher:2006pd,Meszaros:2008zz}
discuss these sources in detail.  Independent of the identities of the
sources, the observed flux of cosmic radiation allows us to make an
estimate of the total number of neutrinos produced by hadrons
accelerated in the universe,\cite{Waxman:1998yy} albeit one somewhat
dependent on assumptions about the opacity and distance distribution
of these sources.\cite{Mannheim:1998wp}  The coming generation of
kilometer-scale neutrino telescopes, such as IceCube and
KM3NeT,\cite{Migneco:2008zz} will be
sensitive to  fluxes significantly below this level, suggesting
that the prospects for detection of VHE neutrino sources in the near
future are good.

\section{The IceCube Detector}

\begin{figure}
  \centerline{\psfig{file=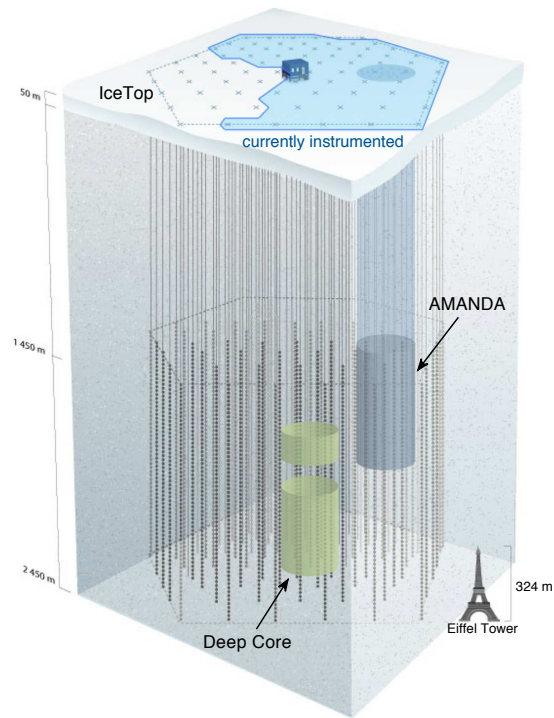,width=2.9in}}
  \vspace*{8pt}
  \caption{Schematic of the IceCube detector.  Each black dot
    represents one DOM.  The surface area corresponding to the strings
    and IceTop stations deployed as of 2009 is shaded blue, and the
    volumes occupied by the AMANDA and Deep Core subarrays are
    indicated.  The Eiffel Tower is shown for
    scale.\protect\label{fig:layout}}
\end{figure}

IceCube consists of a three-dimensional array of photodetectors known
as Digital Optical Modules (DOMs), buried at depths of 1450 m to 2450 m
in the Antarctic ice cap near the South Pole.  The
DOMs are mounted on vertical cables (called ``strings'') of
60 DOMs separated by 17 m.  Eighty such strings will be
deployed on a triangular grid with a horizontal spacing of 125 m
between strings, covering a surface area of approximately 1 km$^2$.
As of the end of the 2008-09 austral summer construction season, a
total of 59 strings have been deployed, as shown in
Fig.~\ref{fig:layout}.  Completion of the full IceCube detector is expected in
February, 2011.

Each DOM houses a 25 cm Hamamatsu photomultiplier tube (PMT) and
associated electronics within a glass pressure vessel.  The onboard
electronics record the PMT signals at a sampling rate of 300 MHz in
the first 400 ns (a slower fADC records late light at a rate of 40
MHz) with an instantaneous dynamic range of approximately 1,000
photoelectrons per 10 ns.  The absolute timing accuracy of each DOM
is better than 3 ns.  The DOMs communicate asynchronously to a
software-based data acquisition (DAQ) system on the surface, which
forms triggers and builds events.\cite{Abbasi:2008ym}

IceCube also includes several specialized subdetectors. An array of
(frozen) water tanks on the surface, known as
IceTop,\cite{Stanev:2009ce} is sensitive to extensive air showers
produced by cosmic rays in the atmosphere. There is one IceTop
station, consisting of two tanks with a diameter of 2 m and an ice
depth of 90 cm, at the top of each IceCube string. Each tank contains
two DOMs, operating at different gains to further extend the dynamic
range. IceTop will be used in conjunction with IceCube to study the
composition of cosmic rays above the ``knee'' in the spectrum around 3
PeV\@. It can also be used for solar physics, as for example with the
measurement of the particle spectra associated with the solar event of
December 13, 2006.\cite{Abbasi:2008vr}

The first-generation South Pole neutrino telescope, AMANDA, is
colocated with IceCube.  Following seven years of operation in its
final configuration, AMANDA was incorporated into IceCube and operated
as a subdetector for two years until being decommissioned in May,
2009.  
The much denser spacing of the AMANDA OMs led to a lower energy
threshold for events in the jointly instrumented
region, but its dissimilar instrumentation
rendered joint operations complex, and its location near the top edge
of IceCube made background rejection difficult.

\subsection{Deep Core}

The IceCube collaboration recently decided to expand the capabilities
of the detector at lower energies with the addition of Deep Core, an
infill array that will be located at the bottom center of IceCube, as
shown in Fig.~\ref{fig:layout}. Deep Core will comprise the seven
innermost standard IceCube strings, as well as six new strings
deployed in a ring of radius 72 m around the central string. The six
new strings will each mount 60 DOMs, 50 of which will be deployed on a
7 m spacing between 2100 m and 2450 m below the surface. The remaining
10 DOMs will be deployed at shallower depths to improve the efficiency
of detection of extremely vertical background muons. With a radius of
125 m and a height of 350 m, the instrumented volume of Deep Core will
include 15 Mton of ice, with expected sensitivity to neutrinos at
energies as low as $\sim$10 GeV.

The new DOMs will be identical to standard IceCube DOMs except that
they will use a new model of PMT developed by Hamamatsu to increase
the quantum efficiency of the photocathode.  Lab tests with assembled
DOMs indicate the sensitivity of the high-QE PMTs is approximately
30\% higher than that of standard DOMs.  The denser DOM spacing and
higher DOM sensitivity combined will increase the collection of
photons in the Deep Core volume by approximately a factor of 10.
Furthermore, the ice at Deep Core depths is significantly more
transparent than that at shallower depths, with optical attenuation
lengths of 40--45 m compared to 20--25 m in the top of the detector.
The significantly improved light collection in Deep Core
translates to much higher sensitivity to relatively dim, low energy
neutrino events.  Additionally, the
bulk of IceCube can be used to detect and veto atmospheric muons
penetrating to Deep Core. The ratio of the atmospheric muon trigger
rate to the atmospheric neutrino trigger rate in IceCube is
approximately $10^6$; initial Monte Carlo studies 
indicate that veto efficiencies on this order are achievable with
relatively good signal efficiency.  

The first of the six new Deep Core strings was successfully deployed
at South Pole in the 2008-09 austral summer.
Deployment of the remaining five new strings, as well as the
standard strings that compose the Deep Core array, is scheduled to be
complete by February 2010.  

\section{Atmospheric Neutrinos}

The flux of atmospheric neutrinos produced in cosmic ray air showers
in the Earth's atmosphere constitutes a background to searches for
astrophysical neutrinos.  They can be distinguished by their much
softer spectrum, $dN/dE_\nu \sim E_\nu^{-3.7}$ at TeV
energies\cite{Volkova:1980sw} instead of the approximately
$E_\nu^{-(2+\delta)}$ generally expected from shock-accelerated
hadronic production.\cite{Learned:2000sw} Also, the flavor ratio
$\nu_e:\nu_\mu:\nu_\tau$ is approximately $0.05:1:0$ at TeV
energies,\cite{Learned:2000sw} as opposed to the approximate flavor
equality expected from astrophysical sources.  At higher energies,
``prompt'' neutrino production in the decay of short-lived charmed
mesons becomes important, leading to a harder spectrum and a flavor
ratio tending to
$1:1:0.1$.\cite{Gondolo:1995fq,Martin:2003us,Enberg:2008te} Relatively
weak limits have been placed on the magnitude of the prompt $\nu_\mu$
flux with AMANDA\cite{Achterberg:2007qp} and better measurements will
be possible with IceCube.

\begin{figure}
  \centerline{
    \begin{minipage}{14pc}
      \psfig{file=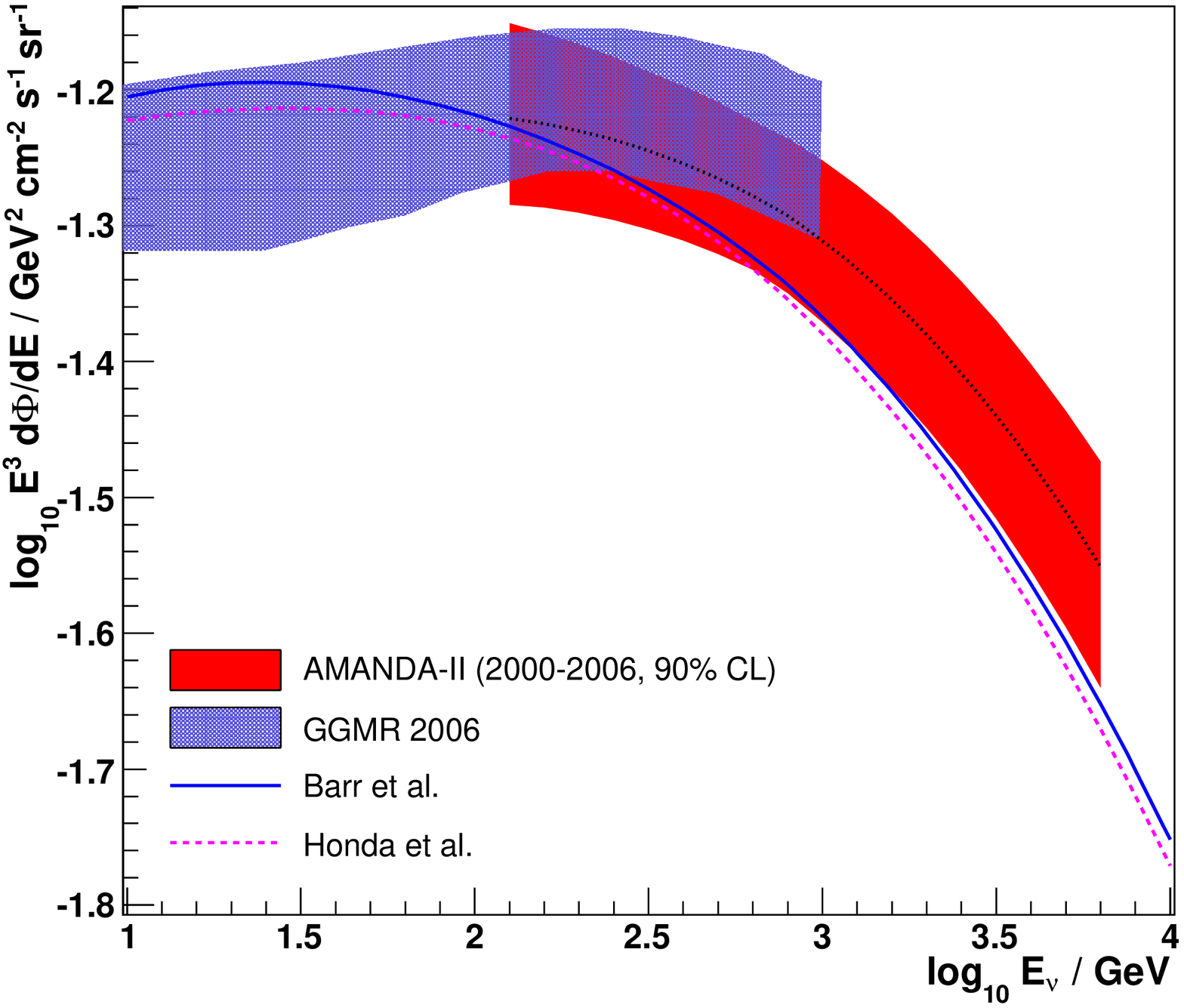,width=2.4in}
    \end{minipage}
    \hspace{1.5pc}%
    \begin{minipage}{14pc}
     \psfig{file=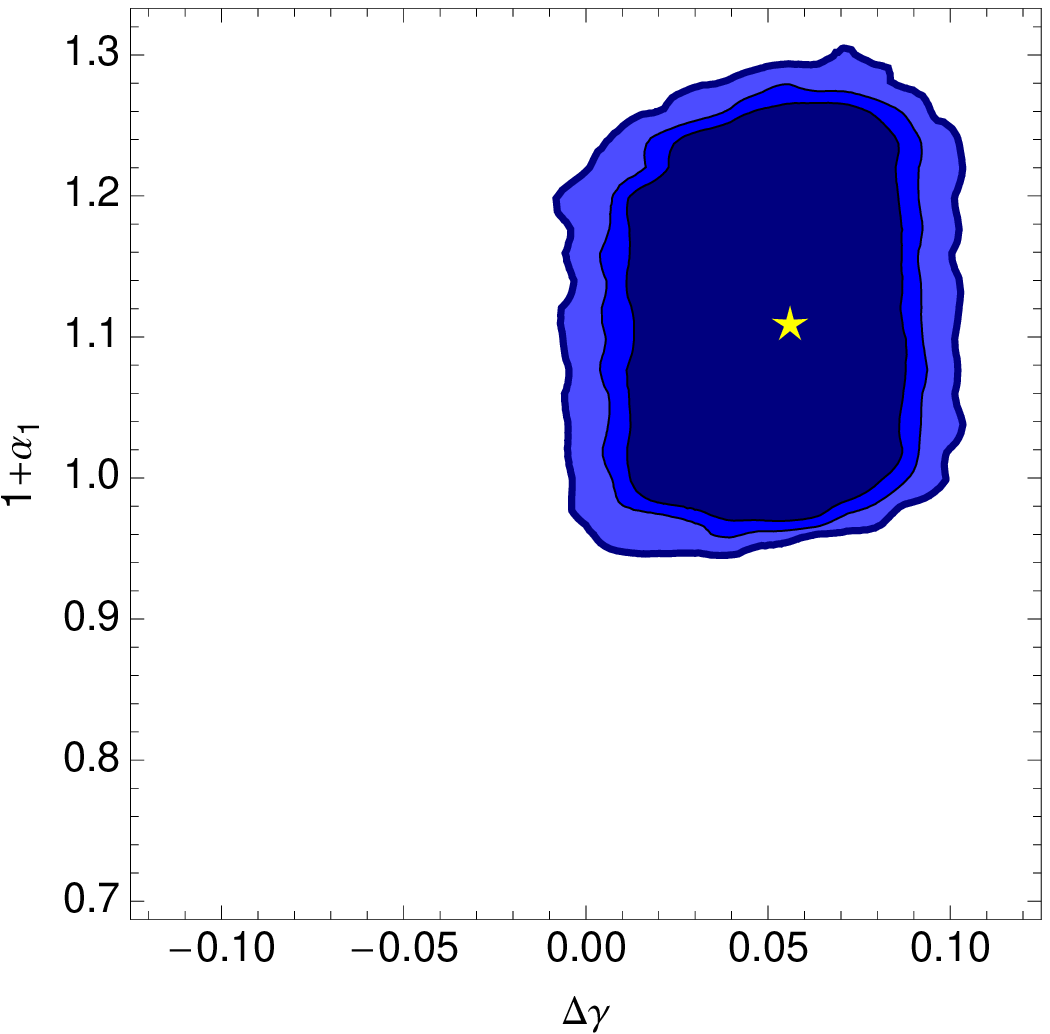,width=2.15in}
     \end{minipage}
   }
   \vspace*{8pt}
   \caption{The atmospheric $\nu_\mu + \bar{\nu}_\mu$ flux observed by
     AMANDA-II.  The 90\%, 95\%, and 99\% C.L. contours and best fit
     point in relative flux normalization and spectral index ($\Delta
     \gamma$) of the flux, relative to the Bartol prediction, are
     shown at right.\protect\label{fig:atmnu}}
\end{figure}

While atmospheric neutrinos constitute a background to astrophysical
neutrinos, their existence also provides a useful tool both for
understanding the response of the IceCube detector and for particle
physics. The atmospheric flux is reasonably well known in the VHE
energy range, with a precision of about 30\% in the muon neutrino rate
at 1 TeV.\cite{Barr:2006it} Two independent calculations of the
atmospheric muon neutrino flux\cite{Barr:2004br,Honda:2006qj} are
shown in Fig.~\ref{fig:atmnu}; note that the deviation between the two
calculations is not a good indicator of the total uncertainty, because
the latter is dominated by uncertainties in external factors used as
inputs by both calculations, such as charged kaon production at high
energies.

A measurement of the atmospheric $\nu_\mu + \bar{\nu}_\mu$ flux using
the full AMANDA-II data set\cite{atmpaper} is shown in
Fig.~\ref{fig:atmnu}, along with a measurement extracted from the
published Super-Kamiokande data set\cite{Ashie:2005ik} by
Gonzalez-Garcia {\it et al.}\cite{GonzalezGarcia:2006ay} The results
begin to suggest a flux slightly (10\%) higher and harder (shift in
spectral index $\Delta \gamma \sim 0.06$) than the central value
calculated by Barr {\it et al.}, but the current AMANDA results are
consistent with the central value at the 90\% to 99\% C.L. in both
parameters. IceCube will rapidly produce a much larger data set, so
improvements may be expected in the near future: the quarter-built
IceCube detector observed roughly as many atmospheric neutrinos
in one year of operation as AMANDA did in seven.

The  large number of atmospheric neutrinos detected by AMANDA can
be used to search for evidence of physics beyond the Standard Model,
such as quantum decoherence or violations of Lorentz
invariance.\cite{atmpaper} These phenomena would appear as
oscillation-like behavior, but at higher energies than standard
neutrino oscillations and with  different characteristic flavor
signatures. The reach of such studies will be greatly improved in the
future with the completion of IceCube and the addition of Deep Core:
IceCube will collect a sample of well over a million atmospheric
neutrino events, with energies ranging from $\sim$10 GeV up to
hundreds of TeV\@. In addition to extending the analyses mentioned
above, such a large neutrino data set should permit observation of
neutrino oscillations, including tau appearance. 
Furthermore, if the $\theta_{13}$ mixing angle is large (near
present limits), a determination of the neutrino mass hierarchy might
be possible, albeit quite difficult, by exploiting matter-enhanced
mixing though the Earth along with asymmetries between the neutrino
and antineutrino interaction cross sections and inelasticity
distributions\cite{Mena:2008rh} at energies around 10 GeV.

\section{Astrophysical Neutrinos}

Searches for individual astrophysical sources of neutrinos
have been conducted separately on both the complete seven year
AMANDA-II data and the one-year data set collected by the 22 IceCube
strings operational in 2007.  Analysis of more recent IceCube data is
underway, and will lead to significant increases in sensitivity.

The AMANDA-II search uses a sky map of 6,595 upward going neutrino
events collected between 2000 and 2006.\cite{Abbasi:2008ih}  The
direction of each of these events was reconstructed using a maximum
likelihood method accounting for the propagation of light in the
ice,\cite{Ahrens:2003fg} with the width of the solution in the
likelihood space used to estimate the angular resolution for each
event and the number of OMs detecting light serving as an energy
estimator.  
A list of 26 energetic galactic and extragalactic objects was defined
{\it a priori} as candidate neutrino sources, including active
galactic nuclei, supernova remnants, and TeV gamma ray sources
identified by Milagro in the Cygnus region\cite{Abdo:2007ad} (now
identified with GeV pulsars or pulsar wind nebulae\cite{Abdo:2009ku}
on the basis of correlations with the Fermi Bright Source
List\cite{Abdo:2009mg}). At the location of each  candidate, a
likelihood ratio was calculated comparing the hypothesis that all
nearby neutrinos were atmospheric to a hypothesis also including a
neutrino source contributing  $\hat{n}_s$ signal events with a
spectral index $\hat{\gamma}$. The significance ($p$-value) of the
likelihood ratio was  assessed by repeatedly randomizing the right
ascension of each event in the data set to create pure background maps
that reproduce any zenith-dependent asymmetries in the detector
response, but with any real neutrino sources smeared out across the
sky. 

\begin{table}[h]
  \tbl{\label{tab:sourcelimits}Selected $p$-values and 90\%
    C. L.~upper limits on $\nu_\mu + \bar{\nu}_\mu$ fluxes
    $E_\nu^2 \, dN/dE \leq \Phi_{90} \times 10^{-12}$ TeV/cm$^2$ s,  from searches 
    for neutrino emission from predefined candidate sources 
    with AMANDA and IceCube-22.  Dashes indicate the best-fit
    number of signal events was zero.}
  {\begin{tabular}{*{7}{lrrrrrr}}
      \toprule            
      Source & decl. [$^\circ$] & r.a. [h] & AMANDA $p$-value &
      IC22 $p$-value & $\Phi_{90}^{\textrm{AM}}$ & $\Phi_{90}^{\textrm{IC22}}$ \cr
      \colrule
      Crab Nebula        & 22.01 &   5.58 & 0.10     & $-$    & 46.4 & 10.35  \cr 
      Geminga             & 17.77 &   6.57 & 0.0086 & $-$    & 63.9 & \hphantom{0}9.67  \cr
      MGRO J2019+37 & 36.83 & 20.32 & 0.077   & 0.25   & 48.4 & 25.23  \cr
      LS I +61 303       & 61.23 &   2.68 & 0.034   & $-$    & 73.7 & 22.00  \cr
      XTE J1118+480   & 48.04 & 11.30 & 0.50     & 0.082 & 25.9 & 40.62 \cr
      Cygnus X-1        & 35.20 & 19.97 & 0.57     & $-$    & 20.0 & 14.60  \cr
      Mrk 421              & 38.21 & 11.07 & 0.82     & $-$    & 12.7 & 14.35  \cr
      Mrk 501              & 39.76 & 16.90 & 0.22     & $-$    & 36.4 & 14.44 \cr
      1ES 1959+650    & 65.15 & 20.00 & 0.44     & 0.071 & 33.8 & 59.00  \cr
      M87                    & 12.39 & 12.51 & 0.43     & $-$    & 22.5 & \hphantom{0}7.91  \cr
      \botrule
    \end{tabular}}
\end{table}

A similar search was carried out using the 2007 IceCube 22-string data
set.\cite{Abbasi:2009iv}  This set, collected from May
2007 to April 2008, contains 5,114 neutrino candidate events detected
in 276 days of live time, consistent with an
expectation of 4600 $\pm$ 1400 atmospheric neutrino events and 400
$\pm$ 200 cosmic ray muons misreconstructed as upgoing neutrino
events.  The source list was expanded in this search to include 28
candidate sources, most of which were identical to the candidates in
the AMANDA-II list. Results for selected sources are shown in
Table~\ref{tab:sourcelimits}, including those with the smallest
$p$-values in each search.  In neither case are these $p$-values
inconsistent with the background hypothesis: one expects to obtain $p
\leq 0.0086$ for at least one of 26 sources in 20\% of signal-free sky
maps, and $p \leq 0.071$ in 66\%.  The 90\%
C. L. upper limits placed on $\nu_\mu$ emission from the sources,
assuming $E^{-2}$ spectra and $1:1:1$ flavor ratios, are
also shown in Table~\ref{tab:sourcelimits}.

\begin{figure}
  \centerline{\psfig{file=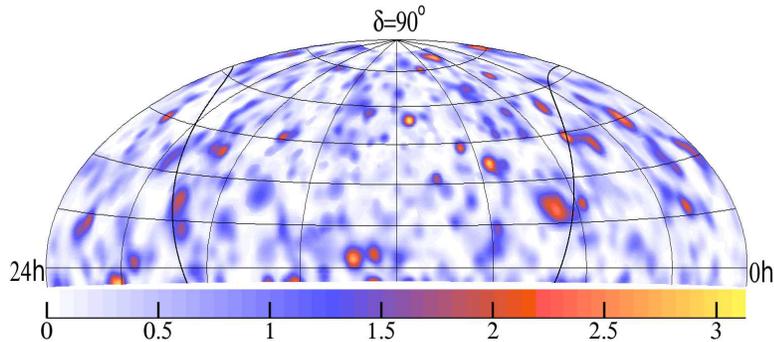,width=0.8\textwidth}}
  \vspace*{8pt}
  \caption{Map of the pre-trials significances (in $\sigma$)
    obtained from an unbinned point source search using the full
    AMANDA-II data
    set.  The most significant point has a significance of
    3.38$\sigma$; 95\% of randomized background skymaps include a
    point at least as significant.\protect\label{fig:AMAskymap}}
\end{figure}

In addition, unbinned all-sky searches were conducted using both data
sets, testing points on $0.25^\circ \times 0.25^\circ$ grids from
declination $-5^\circ$ to $83^\circ$ for the AMANDA-II data set and to
$85^\circ$ for the IceCube-22 data set.  (Higher declinations will be
used to search for dark matter in the Earth's core.)  The
results of the AMANDA-II search are 
shown in Figure~\ref{fig:AMAskymap}.  The most significant point on the
sky, at $\delta = 54^\circ$ and $\alpha = 11.4$ hr, 
had a significance of $3.38\sigma$ before accounting for the large
number of points tested.  In 95\% of randomized maps, a point with at least
$3.38\sigma$ was found, indicating that such a level is 
consistent with statistical fluctuations of the background.

\begin{figure}
  \centerline{\psfig{file=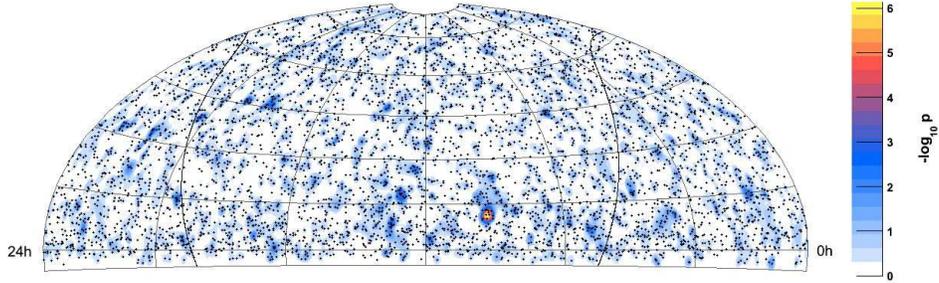,width=\textwidth}}
  \vspace*{8pt}
  \caption{Map of the pre-trials $p$-values obtained from an unbinned
    point source search using the 22-string 2007 IceCube data set.
    Dots indicate the individual events used to calculate the
    significances.  The probability of observing in signal-free maps a
    point at least as significant as the brightest spot in this map is
    estimated to be 1.34\%, insufficient to reject the background
    hypothesis.\protect\label{fig:IC22skymap}}
\end{figure}

The results of the all-sky search using IceCube data are shown in
Fig.~\ref{fig:IC22skymap}. The most significant deviation from the
background expectation, with $\hat{n}_s = 7.7$ events, occured at
153.4$^\circ$ right ascension, 11.4$^\circ$ declination. The
pre-trials significance of this deviation is estimated to be $7 \times
10^{-7}$, but more significant deviations were observed in 67 out of
10,000 scrambled background maps. Accounting for the
parallel search for emission from sources selected {\it a priori} was
also performed on the data set, as described above, the final
$p$-value of the result is estimated to be 1.34\%. This is not
sufficient to reject the background-only null hypothesis.

\begin{figure}
  \centerline{\psfig{file=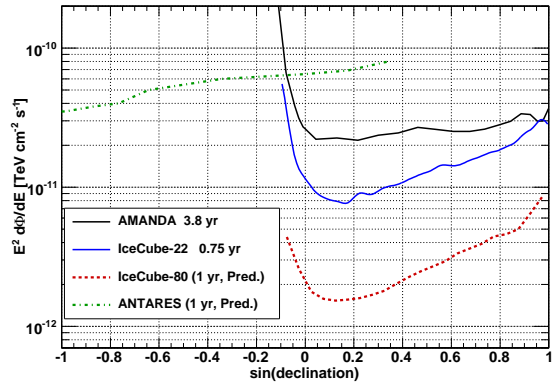,width=3in}}
  \vspace*{8pt}
  \caption{Average 90\% C.L. upper limit on the $\nu_\mu +
    \bar{\nu}_\mu$ flux (assuming flavor equality) from a
    point source as a function of declination, compared to
     predictions for IceCube and
    ANTARES.\protect\label{fig:ptsrcsens}}
\end{figure}

The sensitivities of the two all-sky searches are compared in
Fig.~\ref{fig:ptsrcsens}, for $\nu_\mu$ fluxes $dN/dE \sim E^{-2}$.
Predicted sensitivities for one year of operation with the full
IceCube and with ANTARES\cite{AguilarSanchez:2007hh} are also
shown. One year of operation with only one quarter of the full IceCube
array already provides a sensitivity significantly surpassing that of
the seven-year AMANDA-II data set.  This sensitivity will increase
rapidly as IceCube construction is completed, so if the ``hot spot''
seen in the IC22 sky map were indicative of a neutrino source rather
than simply a background fluctuation, the source would be definitively
detected in the very near future. 

\subsection{Gamma ray bursts}

Searches for neutrino emission in coincidence
with gamma ray bursts (GRBs) observed via their electromagnetic
emission have also been conducted, with both AMANDA and IceCube.
These analyses benefit from drastically lower background rates, due to
the external information regarding the time and direction of the
bursts.  Nevertheless, the neutrino event rates are low (typically
$\ll$ 1) for individual bursts, so the searches are normally conducted
by `stacking' all known bursts and looking for neutrinos in
conjunction with the ensemble as a whole.

\begin{figure}
  \centerline{
    \begin{minipage}{0.42\textwidth}
      \psfig{file=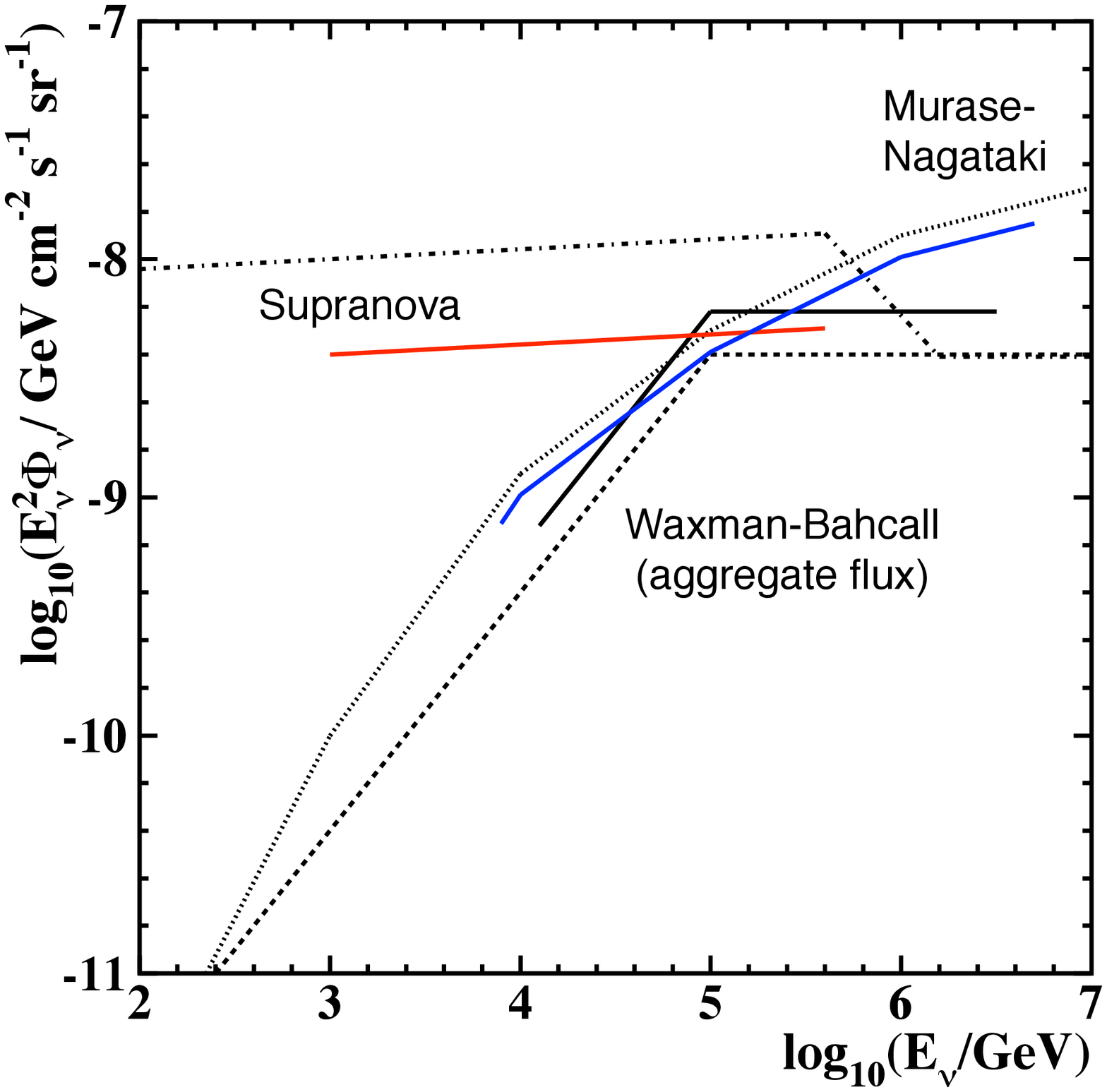,width=\textwidth}
    \end{minipage}
    \hspace{1.5pc}%
    \begin{minipage}{0.56\textwidth}
     \psfig{file=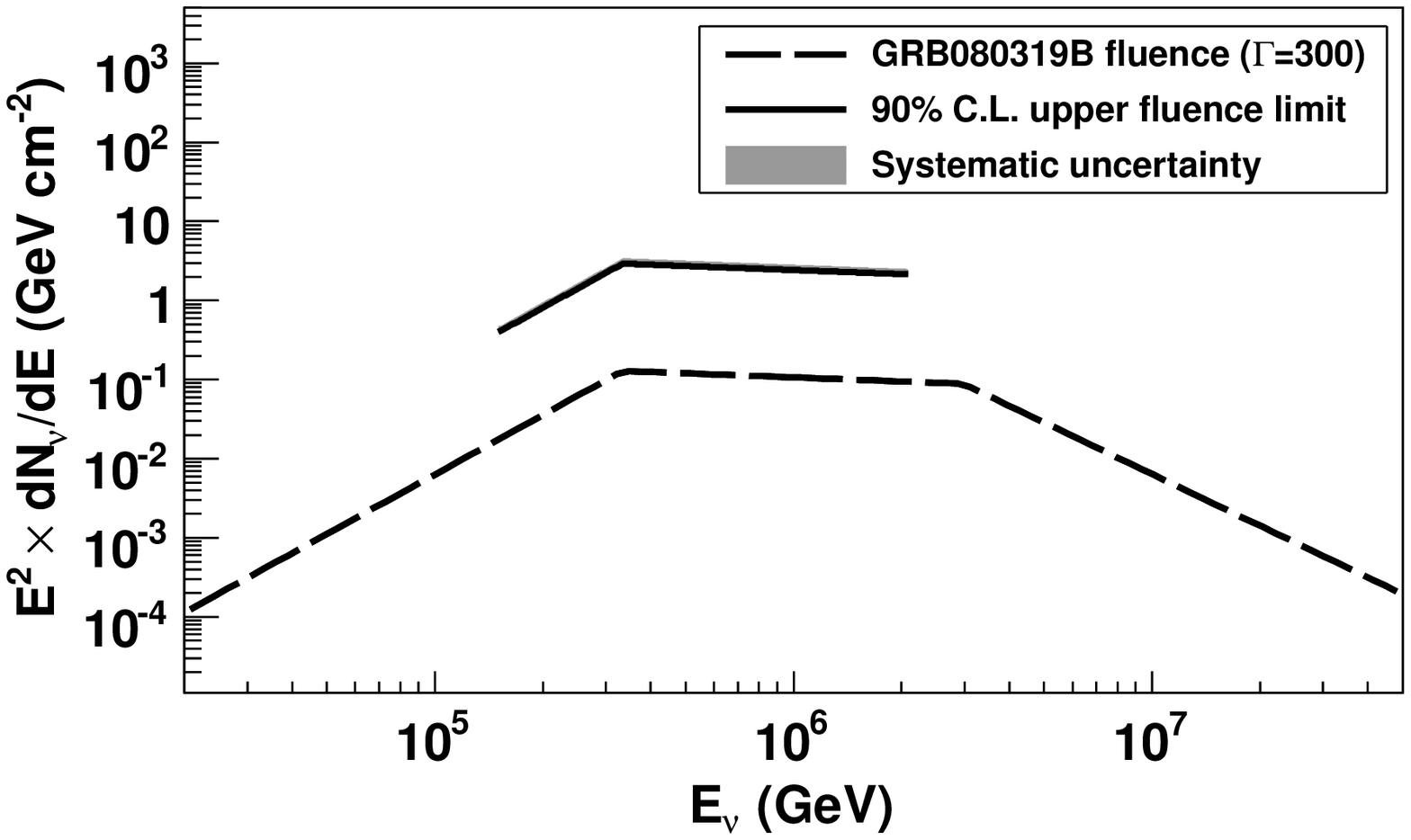,width=\textwidth}
     \end{minipage}
   }
   \vspace*{8pt}
   \caption{{\it Left:} Integral limits at 90\% C. L. on several $\nu_\mu + \bar{\nu}_\mu$
      flux models (details in text).  The energy ranges indicated
      contain 90\% of the expected flux.  The flux limits are for the full sky ($4\pi$ sr),
      although only bursts from the Northern Hemisphere were used in
      the analysis.  {\it Right:}  Predicted $\nu_\mu + \bar{\nu}_\mu$
      flux from for GRB 080319B
      using the Waxman-Bahcall model with observed burst
      parameters and assuming $\Gamma_\textrm{jet} = 300$, compared to
      the 90\% C. L. upper limit.\protect\label{fig:grb}}
\end{figure}

The results of searches for emission from 419 GRBs observed in the
Northern Hemisphere during stable AMANDA operations between 1997 and
2003\cite{Achterberg:2007nx} is shown in
Fig.~\ref{fig:grb} (left). Three theoretical models are shown: the
Waxman-Bahcall\cite{Waxman:2002wp} (divided by 2 to account for
neutrino oscillations) and Murase-Nagataki\cite{Murase:2005hy}
calculations based on the assumption that GRBs are the sources of the
ultrahigh-energy cosmic rays, and the `supranova' model of Razzaque
{\it et al.},\cite{Razzaque:2002kb} under the assumption that all GRBs
are preceded by supernovae which produce ideal circumburst
environments for neutrino production. For the latter two models, only
the flux from long-duration GRBs is included; both long and short are
included for the Waxman-Bahcall model.

Because limits are placed on the integrated flux predicted by the
models, our constraints are given in terms of ``model rejection
factors'' (MRFs), essentially the scaling factor at which the model would be
just ruled out at 90\% confidence.\cite{Hill:2002nv}  MRFs less than 1
indicate that the model could be scaled down and still be excluded;
i.e., that the model is excluded at the stated confidence level.  The
MRFs for the three models are 1.36 for the Waxman-Bahcall model, 0.92
for the Murase-Nagataki parameter set A, and 0.45 for the supranova model under the
assumptions mentioned above.  A comparable search using IceCube
observations is in progress.

A dedicated search was also conducted for neutrinos from the
``naked-eye'' GRB 080319B.\cite{Abbasi:2009kq}  Although 40 strings were in the ice at the
time of this burst, the 18 new strings deployed in the 2007-08 austral
summer had not yet been commissioned.  More unfortunately, the burst
occured during a maintenance period related to the transition to
40-string operations, with only nine strings taking data.
Nonetheless, the extreme nature of the burst leads to a non-negligible
predicted event rate in the 9-string detector.  The limit placed on
the Waxman-Bahcall flux, with parameters chosen according to those
measured for GRB 080319B, is shown in Fig.~\ref{fig:grb} (right) for an
assumed jet Lorentz factor $\Gamma_\textrm{jet} = 300$.  For
this choice of $\Gamma_\textrm{jet}$, 0.12 muon events are predicted,
leading to an MRF of 22.7.  Higher values of $\Gamma_\textrm{jet}$
lead to lower predicted event rates.  Although the number of events
expected for the 9-string configuration then operational is low, a
similar burst observed with the full IceCube detector would be
expected to produce $\mathcal{O}(1)$ event.

\subsection{Diffuse fluxes}

IceCube could also detect diffuse fluxes of high energy neutrinos, for
example from a class of sources too dim to resolve
individually. At relatively low energies such a flux would be
indistinguishable from the atmospheric neutrino flux, but the soft
spectrum of the conventional atmospheric neutrinos means that a
diffuse astrophysical flux could be detected as a harder spectral
component of the total diffuse flux.

\begin{figure}
  \centerline{\psfig{file=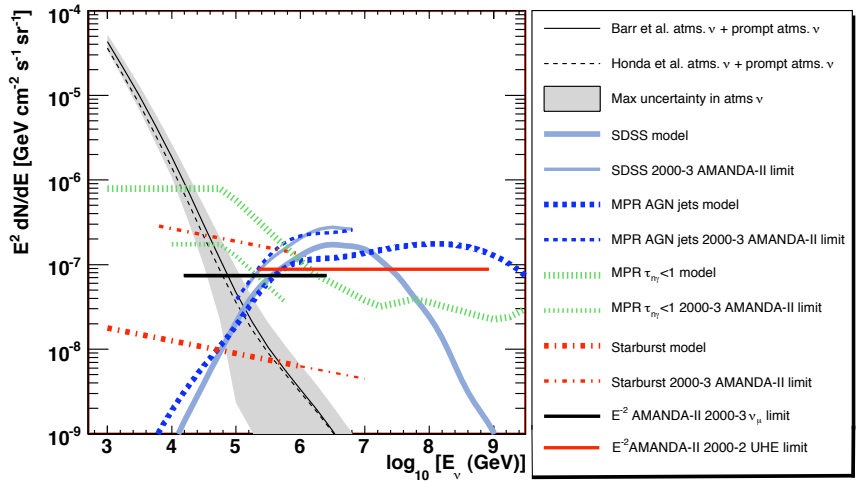,width=4.5in}}
  \vspace*{8pt}
  \caption{AMANDA diffuse and ultrahigh-energy (UHE) integral limits
    on hypothetical $\nu_\mu + \bar{\nu}_\mu$ fluxes, compared to
    several diffuse flux predictions (thick lines, details in text)
    and the limits on those predictions (thin lines).  The energy
    range for each limit is that within which 90\% of detected neutrino
    events would fall.\protect\label{fig:diffuseAstro}}
\end{figure}

Two independent searches for diffuse fluxes were undertaken using
AMANDA data. The first\cite{Achterberg:2007qp} was closely related to
the standard muon neutrino analysis used to search for point sources
of neutrinos, with cuts optimized for higher energy neutrinos and with
a cut placed on the reconstructed energy of the neutrino candidate
events. This search used data from 2000--2003, and the results are
shown in Fig.~\ref{fig:diffuseAstro}. As a benchmark, the limit placed
on a hypothetical diffuse $\nu_\mu$ flux $dN/dE = \Phi_0 E^{-2}$ at
90\% C. L. is $\Phi_0 \geq 7.4 \times 10^{-8}$ GeV cm$^{-2}$ s$^{-1}$
sr$^{-1}$. For such a flux, 90\% of the signal neutrinos would have
had energies between 16 TeV and 2.5 PeV, which are the bounds of limit
shown in Fig.~\ref{fig:diffuseAstro}. The limit is placed on the
integrated flux, and so cannot be directly compared to specific
theoretical models; instead, the predicted signal for each model must
be simulated to find the model rejection factor (MRF). Limits on several
specific models are also shown (as thin lines parallel to the
predicted fluxes) in Fig.~\ref{fig:diffuseAstro}, and appear in
Table~\ref{tab:diffuse}. In addition to astrophysical models, a flux
corresponding to the upper bound on generic optically thin
($\tau_{n\gamma} < 1$) pion photoproduction
sites\cite{Mannheim:1998wp} was tested. The upper limit from the
analysis is a factor of 0.22 of the MPR bound over the region from 10
TeV to 630 TeV.

A second analysis based on data from 2000--2002 exploited
idiosyncracies of the hardware response to the extremely bright events
produced by ultrahigh energy (UHE) neutrinos,
such as afterpulsing in the PMTs, to extend the range of the detector
to much higher energies.\cite{Ackermann:2007km} The limit for the
$E^{-2}$ benchmark $\nu_\mu$ flux is $9.0 \times 10^{-8}$ GeV
cm$^{-2}$ s$^{-1}$ sr$^{-1}$ at 90\% C. L. (assuming a $1:1:1$ flavor
ratio), over a range from $2 \times 10^5$ GeV to $10^9$ GeV, as shown
in Fig.~\ref{fig:diffuseAstro}. It should be noted that the analysis
was sensitive to all flavors of neutrinos; for a $E^{-2}$ spectrum
with flavor equality, the flavor ratio of the detected events would
have been approximately $2:2:1$.  Limits on specific theoretical models
need to be calculated separately for this analysis and are omitted
from Fig.~\ref{fig:diffuseAstro} for clarity but are shown in
Table~\ref{tab:diffuse}.  

\begin{table}[h]
  \tbl{\label{tab:diffuse}Limits on several theoretical models of diffuse
    muon neutrino fluxes from the two analyses.  The number of events that
    would have been detected $n_{sig}$ and the ``model rejection
    factor'' (MRF), the ratio of the upper limit to
    the predicted flux, are shown.  MRFs less than 1 indicate that the
    model is excluded at the 90\% C. L.}
  {\begin{tabular}{@{}lrrrrl@{}} \toprule
      Source & $n_{sig}^\textrm{HE}$ & MRF$^\textrm{HE}$ &
      $n_{sig}^\textrm{UHE}$ & MRF$^\textrm{UHE}$ & Model \\ 
      \colrule
      Active Galactic Nuclei          
      & 1.7   & 1.6   &   1.8 &   2.9 & Stecker\cite{Stecker:2005hn} \\
      & 1.4   & 2.0   &   5.9 &   0.9 & MPR\cite{Mannheim:1998wp} \\
      &         &         &   8.8 &   0.6 & Halzen \& Zas\cite{Halzen:1997hw} \\
      &         &         & 20.6 &   0.3 & Protheroe\cite{Protheroe:1996uu} \\
      &         &         &   0.3 & 18.0 & Mannheim\cite{Mannheim:1995mm} RL A \\
      &         &         &   4.5 &   1.2 & Mannheim\cite{Mannheim:1995mm} RL B \\
      Starburst Galaxies
      & 1.1   & 21.1 &         &         & Loeb \& Waxman\cite{Loeb:2006dk} \\ 
      Prompt Atmospheric $\nu_\mu$
      & 0.4   & 60.3 &         &         & MRS GBW\cite{Martin:2003us} \\
      & 4.7   &   5.2 &         &         & Naumov RQPM\cite{Naumov:2002dm} \\
      & 16.1 &   1.5 &         &         & Zas et al.\cite{Zas:1992ci} Charm C \\
      & 26.2 & 0.95 &         &         & Zas et al.\cite{Zas:1992ci}  Charm D \\
      \botrule
\end{tabular}}
\end{table}


\section{Dark Matter}

Large volume neutrino telescopes such as IceCube can search for
indirect evidence of dark matter, such as supersymmetric neutralinos,
that could accumulate in the gravitational wells of the Earth and Sun
and annihilate to produce neutrinos.  These searches
are complementary to those conducted by direct detection experiments,
because the latter generally rely on coherent scattering of the WIMP from the
ensemble of nucleons in heavy nuclei. Direct detection experiments
thus constrain primarily the spin-independent neutralino-nucleon
scattering cross section $\sigma_\textrm{SI}$. Neutralino capture in
the Sun, which is made primarily of light nuclei, allows us to probe
models where the coupling is primarily
spin-dependent.\cite{Halzen:2005ar} It should also be noted that
direct and indirect searches probe different epochs of the history of
the solar system and different parts of the WIMP velocity
distribution.

\begin{figure}
  \begin{minipage}{0.49\textwidth}
    \centerline{\psfig{file=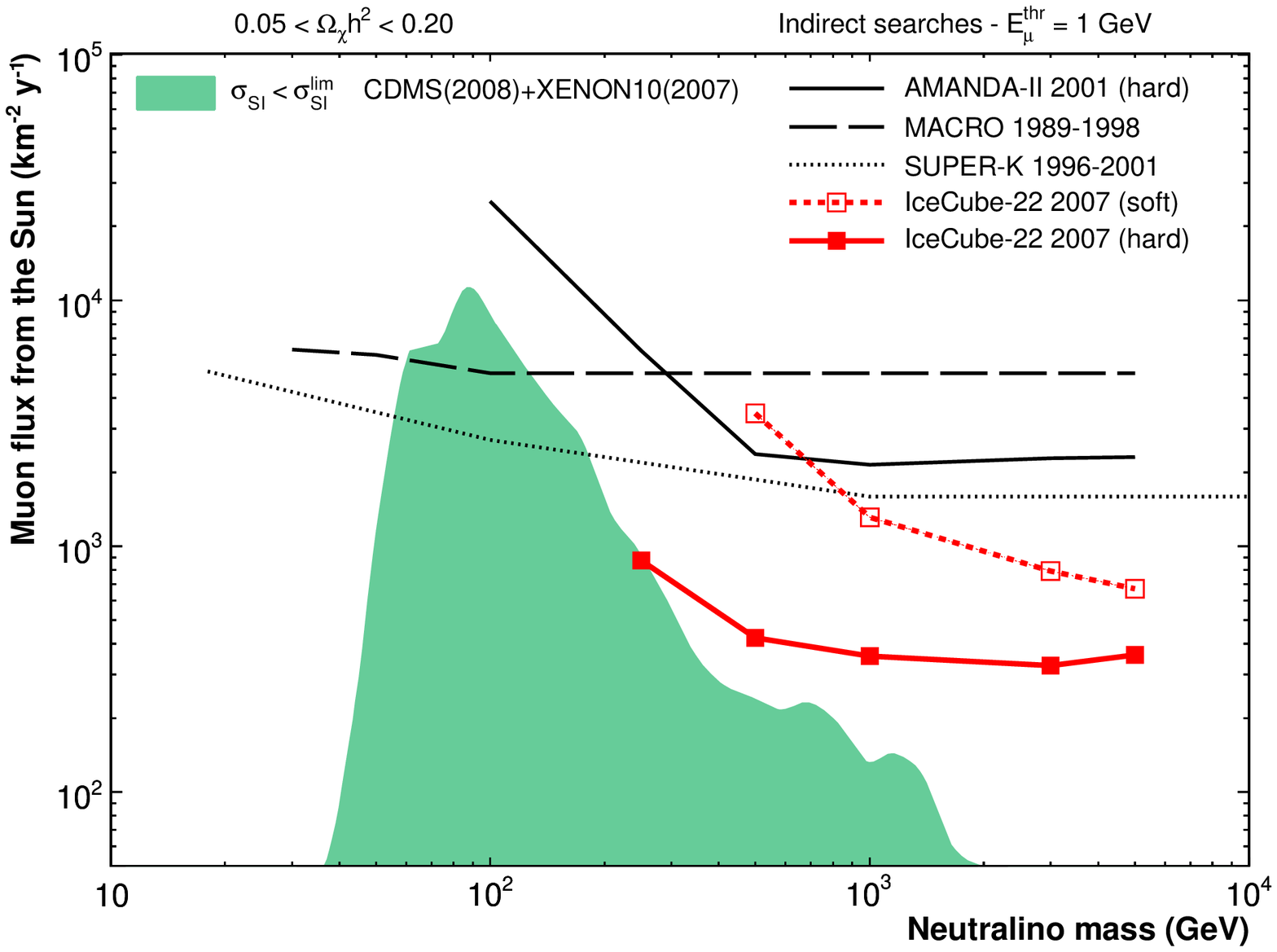,width=\textwidth}}
  \end{minipage}
  \hspace{2pt}
  \begin{minipage}{0.49\textwidth}
    \centerline{\psfig{file=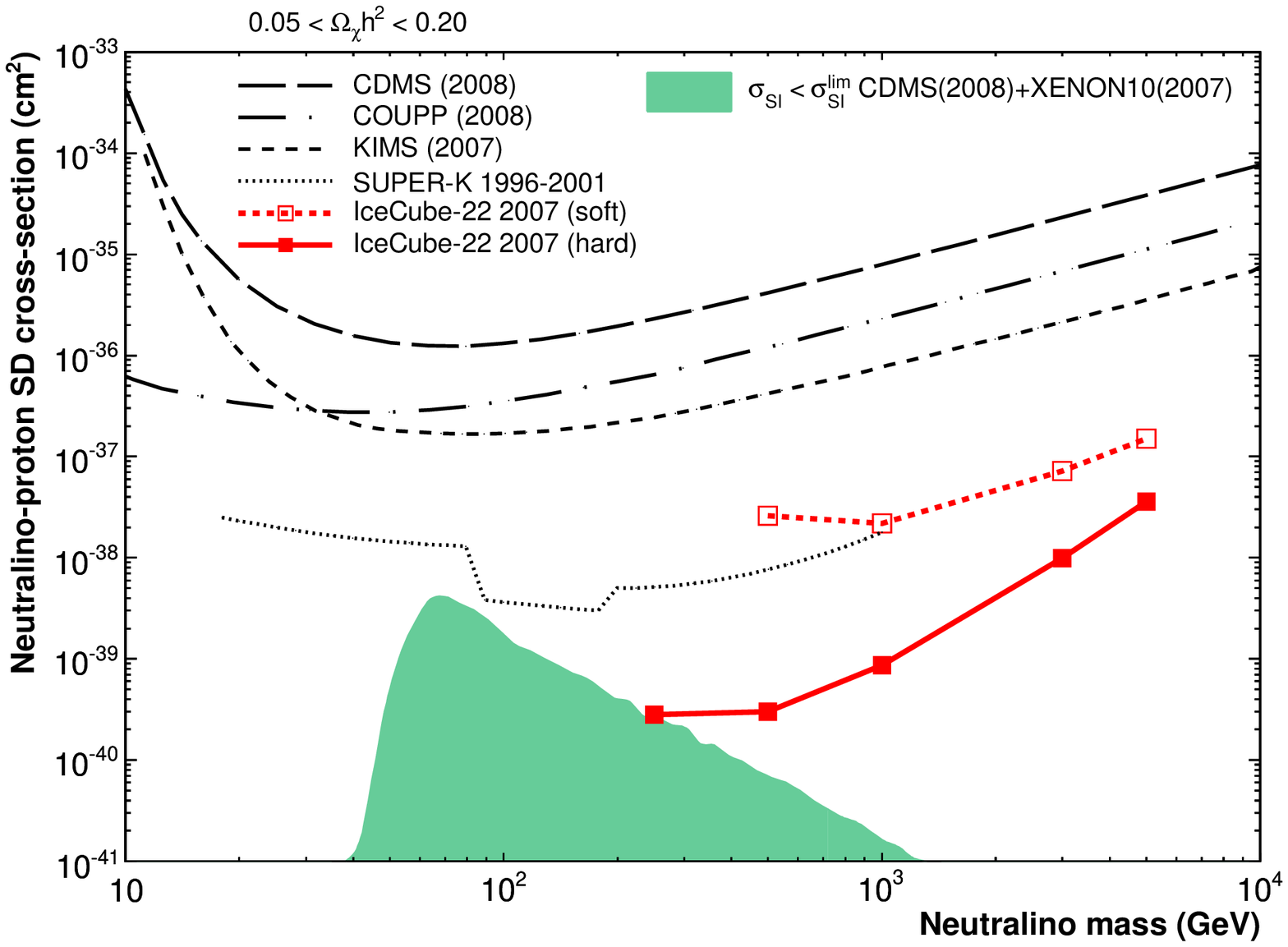,width=\textwidth}}
  \end{minipage}
  \vspace*{8pt}
  \caption{ {\it Left:} Limits on the flux of neutrino-induced muons
    due to neutralino annihilation in the Sun, normalized to a 1 GeV
    muon threshold with assumed hard ($\chi \chi \rightarrow W^+W^-$) and
    soft ($b\bar{b}$) neutrino spectra. The shaded region indicated
    the allowed MSSM parameter space. {\it Right:} Limits on the
    spin-dependent neutralino-proton cross section inferred based on
    the muon limits.\protect\label{fig:wimpLimit}}
\end{figure}

Figure~\ref{fig:wimpLimit} (left) shows the limit on the flux
of muons produced by neutrinos generated in neutralino annihilations
in the sun with the 22-string IceCube detector,\cite{Abbasi:2009uz}
compared to those set by other neutrino
detectors.\cite{Ambrosio:1998qj,Desai:2004pq,Ackermann:2005fr} This
analysis used 104 days of live time from June to September 2007, while
the Sun was below the horizon as viewed from the South Pole. This
limit includes a full simulation of WIMP annihilation in the Sun and
neutrino transport through the Sun and Earth\cite{Blennow:2007tw} as
well as of the IceCube detector, and for comparison between
experiments the flux is quoted above an arbitrary threshold of 1
GeV\@. In Figure~\ref{fig:wimpLimit} (right), the limit on the WIMP
annihilation rate shown in Fig.~\ref{fig:wimpLimit} (left) is
converted to a limit on the spin-dependent scattering cross section
$\sigma_\textrm{SD}$, using the conservative assumption that
$\sigma_\textrm{SI} = 0$. The limits on $\sigma_\textrm{SD}$ from
direct detection
experiments\cite{Angle:2007uj,Ahmed:2008eu,Lee.:2007qn,Behnke:2008zza}
and Super-K\cite{Desai:2004pq} are also shown. In both figures, the
shaded area represents the muon flux or $\sigma_\textrm{SD}$ predicted
by MSSM models not excluded by direct detection experiments on the
basis of their limits on $\sigma_\textrm{SI}$.  It should be noted
that most of this model space includes relatively weak
spin-independent coupling, so that very significant improvements (more
than 3 orders of magnitude) in sensitivity would be required for
direct detection experiments to significantly constrain the remaining
models.  An analysis of the sensitivity to other types of dark matter,
such as Kaluza-Klein particles, is in progress.

The future reach of dark matter searches will be greatly
enhanced by the completion of IceCube and the addition of the Deep
Core array, which will substantially improve IceCube's sensitivity to
neutralino masses below a few hundred GeV.

\section{Outlook}

Construction of the IceCube detector is proceeding very well, with
completion of the array expected in 2011.  IceCube will be augmented
with the Deep Core array, to be completed in 2010, which will
significantly extend its capabilities at energies as low as 10 GeV.
Initial results from the partially built detector, including only one
quarter of the final array, are already providing sensitivities beyond
those of the complete seven-year AMANDA-II data set,
and this sensitivity will expand rapidly as construction progresses.
Within a few years the sensitivity of IceCube will be sufficient to
detect astrophysical neutrino fluxes at the levels expected on
theoretical grounds to be produced by the sources of the cosmic rays.
In addition, IceCube and Deep Core will permit indirect searches for dark
matter well beyond existing limits, and studies of atmospheric
neutrinos with unprecedented statistics.

\section*{Acknowledgments}

The author is grateful for the hospitality of the Aspen Institute and
for valuable discussions with many participants in the 2009 neutrino physics
workshop during the preparation of this review.  This work was supported by NSF
grant PHY-0554868.

\end{document}